\documentclass [12pt,a4paper]{article}
\usepackage{bbm}
\usepackage{amsfonts}
\usepackage{mathrsfs}
\usepackage {amssymb}
\usepackage {amsmath}
\usepackage{amsthm}
\usepackage{latexsym}
\usepackage {amssymb}
\usepackage {amsmath}
\usepackage{color}
\usepackage{enumerate}
\theoremstyle{definition}

\newcommand{\F}{\mathbb{F}}
\newcommand{\Z}{\mathbb{Z}}

\makeatletter
 \@addtoreset{equation}{section}
 
\makeatother

\usepackage{lastpage}
\usepackage{epsfig}

\newcommand{\pf}{{\bf Proof. \ }}

\begin{document}
\begin{center}
\textbf{\Large{Cyclotomic Construction of Strong External Difference Families in Finite Fields} }\footnote {The work of J. Wen and F. Fu was supported by the National Key Basic Research
Program of China under Grant 2013CB834204, and the NSFC under Grant 61571243, 61171082.
The work of M. Yang was supported by the NSFC under Grant 61379139, 11526215. The work of K. Feng was supported by the NSFC under Grant 11571107 and 11471178.\\
J. Wen and F. Fu are with the Chern Institute of Mathematics, Nankai University, Tianjin 300071, China(e-mail:jjwen@mail.nankai.edu.cn, fwfu@nankai.edu.cn)\\
 M. Yang is with the State Key Laboratory of Information
Security, Institute of Information Engineering, Chinese Academy
of Sciences, Beijing 100193, China(e-mail:yangminghui6688@163.com)\\
K. Feng is with the Department of Mathematical Sciences, Tsinghua University, Beijing, 100084, China(e-mail: kfeng@math.tsinghua.edu.cn)}

\end{center}

\begin{center}
\small Jiejing Wen, Minghui Yang, Fangwei Fu,  Keqin Feng
\end{center}


\noindent\textbf{Abstract} Strong external difference family (SEDF) and its generalizations GSEDF, BGSEDF in a finite abelian group $G$ are combinatorial designs raised by Paterson and Stinson [7] in 2016 and have applications in communication theory to construct optimal strong algebraic manipulation detection codes. In this paper we firstly present some general constructions of these combinatorial designs by using difference sets and partial difference sets in $G$. Then, as applications of the general constructions, we construct series of SEDF, GSEDF and BGSEDF in finite fields by using cyclotomic classes.

\noindent\textbf{Key Words} strong external difference family, difference set, partial difference set, cyclotomic class, cyclotomic number, finite field, strong algebraic manipulation detection code

\section{Introduction}

\ \ \ \ Let $(G,+)$ be an abelian group. For subsets $D_{1}$ and $D_{2}$ of $G$, $|D_{1}|, |D_{2}|\geq 1$,
we define the multiset
$$\triangle(D_{1}, D_{2})=\{a_1-a_2: a_1\in {D_1}, a_2\in {D_2}\}$$
A subset $D$ of $G$ is called an $(n,k,\lambda)$-DS (difference set) in $G$ if $|G|=n, |D|=k$ and
\begin{equation}
\Delta(D, D)=k\{0\}+\lambda(G-\{0\})
\end{equation}
where we use the notation in group ring $\mathbb{Z}[G]$ to express the multisets in both sides of (1.1). Namely, the equality (1.1) means that the multiplicity of any nonzero element $g\in G$ in
 $\Delta(D, D)$ is $\lambda$ (and the multiplicity of zero in $\Delta(D, D)$ should be $k=|D|$).

 Suppose that $0\notin D$, $D$ is called $(n,k,\lambda,\mu)$-PDS (partial difference set) in $G$ if
 \begin{equation}
\Delta(D, D)=k\{0\}+\lambda D+\mu(G-D-\{0\})
\end{equation}

Difference set and its generalizations are important objects in combinatorial design theory. Some generalizations and variations come from their applications in other areas, as statistics and digit communication. In this paper we focus on the following variation raised in [7].

$\mathbf{Definition \ 1.1}$ Let $G$ be an abelian group, $|G|=n,$ $m\geq2$, $A_i$ $(1\leq i\leq m)$
be subsets of $G$ and $|A_i|=k\geq 1$ $(1\leq i\leq m)$. The family $\{A_1,\ldots, A_m\}$ is called an $(n,m,k,\lambda)$-SEDF (strong external difference family) in $G$ if for each $i$, $1\leq i\leq m$,
\begin{equation}
\sum^{m}_{j=1\atop j\neq i }\Delta(A_{i},A_{j})=\lambda(G-\{0\})
\end{equation}

It is easy to see that the equality (1.3) implies that $(m-1)k^{2}=\lambda(n-1)$ and $A_{1},\ldots,A_{m}$ should be pairwise disjoint. Particularly, $\{A_{1},A_{2}\}$ is an $(n,2,k,\lambda)$-SEDF in $G$ if and only if $|A_{1}|=|A_{2}|=k$ and $\Delta\{A_{1},A_{2}\}=\lambda(G-\{0\})$.

More general, a family $\{A_1,\ldots, A_m\}$ $(A_{i}\subseteq G)$ is called an $(n,m;k_{1},\ldots,k_{m};\lambda_{1},\ldots,\lambda_{m})$-GSEDF (generalized SEDF) in G if $|A_{i}|=k_{i}\geq1$ $(1\leq i\leq m)$ and for each $i$,
\begin{equation}
\sum^{m}_{j=1\atop j\neq i }\Delta(A_{i},A_{j})=\lambda_{i}(G-\{0\})
\end{equation}

Let $k=k_{1}+\cdots+k_{m}$. The equality (1.4) implies that $k_{i}(k-k_{i})=\lambda_{i}(n-1)$ $(1\leq i\leq m)$ and $A_{1},\ldots,A_{m}$ should be pairwise disjoint.

Finally, even more general, $\{A_{1},\ldots,A_{m}\}$ is called $\{n,m;k_{1},\ldots,k_{m};\lambda_{1},\ldots,\lambda_{m}\}$-BGSEDF (Bounded GSEDF), if $|A_{i}|=k_{i}\geq 1$ $(1\leq i\leq m)$ and for each $i$,
$$\sum^{m}_{j=1\atop j\neq i }\Delta(A_{i},A_{j})\leq \lambda_{i}(G-\{0\})$$
where for two elements $\alpha=\sum_{g\in G}a_{g}\{g\}$ and  $\beta=\sum_{g\in G}b_{g}\{g\}$ in group ring $\Z[G],\alpha\leq\beta$ means that $a_{g}\leq b_{g}$ for each $g\in G$.

These combinatorial designs have applications in communication theory to construct optimal strong algebraic manipulation detection (AMD) codes. Roughly speaking, there are two types of optimal strong AMD codes: $R$-optimal and $G$-optimal, which reach two kinds of lower bounds of maximum success probabilities respectively. From $(n,m;k_{1},\ldots,k_{m};\lambda_{1},\ldots,\lambda_{m})$-GSEDF's we can get
$R$-optimal strong AMD codes ([7], Theorem 4.10). If $k_1=\cdots =k_m=k$ and $\lambda_1=\cdots= \lambda_m=\lambda$, then from $(n,m;k,\lambda)$-SEDF's we can get ``k-uniform" R-optimal strong AMD codes.
On the other hand, $G$-optimal strong AMD codes are equivalent to $(n,m;k_{1},\ldots,k_{m};1,\ldots,1)$-BGSEDF's. For the definition of strong AMD code and the relationship
between strong ADM codes and the designs GSEDF, BGSEDF we refer to [7].

In recent two years, several existence and non-existence results on $(n,m;k,\lambda)$-SEDF have been proved
[7,6,4,9]. For any abelian group $G=\{g_1,\ldots,g_n\}$, we have trivial SEDF $\{g_1\}$, $\{g_2\}$,$\ldots,$
$\{g_n\}$ with parameters $(n,n;1,1)$. From now on we concern on nontrivial case. Several series of SEDF's
with $m=2$ have been constructed ([7, 6, 4]). On the other hand, Martin and Stinson [6] proved that there is no (nontrivial) SEDF with $m=3, 4$. More nonexistence results are given by Huczynska and Paterson [4] and Bao et al. [2]. The open problem on existence of SEDF with $m\geq 3$ is raised in [7] and continuously asked in [6], [4] and [2] for $m\geq 5$. Wen et al. [10] present an example of (243, 11; 22, 20)-SEDF by using the cyclotomic classes of order 11 in finite field $\mathbb{F}_q$ $(q=3^5=243).$ At this moment, the problem on
existence of SEDF with $m\geq 5$ and $m \neq 11$ is still open.

 The paper is organized as following. Firstly we show several general constructions of SEDF and GSEDF in Section 2 by using DS and PDS. As
applications of these general results, we construct series of SEDF with $m=2$, GSEDF and BGSEDF with $\lambda_1=\cdots=\lambda_m=1$ in finite field $\mathbb{F}_q$ by using cyclotomic classes in $\mathbb{F}_q$
(Section 3). The last section is conclusion.

\section{General Constructions of SEDF and GSEDF}

We start with a result in [7] which says that if $A_1,\ldots,A_m$ is a partition of a finite abelian group $G$, then $\{A_1,\ldots,A_m\}$ is a GSEDF in $G$ if and only if each $A_i$ is a DS in $G$.

$\mathbf{Theoem\ 2.1}$ ([7], Theorem 2.4) Suppose that $A_1,\ldots,A_m$ $(m\geq2)$ is a partition of a finite abelian group $(G,+)$. Then $\{A_1,\ldots,A_m\}$ is a GSEDF in $G$ with parameters $(n,m;k_{1},\ldots,k_{m};\lambda_1,\ldots,\lambda_m)$ if and only if each $A_i$ $(1\leq i\leq m)$ is a
DS in $G$ with parameters $(n,k_i,k_i-\lambda_i)$.

For each $g\in G$, $\{g\}$ is trivial $(n,1,0)$-DS in $G$. From Theorem 2.1, we get

$\mathbf{Corollary\ 2.2}$ Let $A_{1},\ldots, A_{m}$ be pairwise disjoint subsets of an abelian group $G$, $|G|=n$, $|A_{i}|=k_{i}\geq1$ $(1\leq i\leq m)$. Let $G-\sum_{i=1}^{m}A_{i}=\{g_{1},\ldots,g_{r}\}$ $(r=n-\sum_{i=1}^{m}k_{i})$. If each $A_{i}$ is an $(n,k_{i},\lambda_{i}^{'})$-DS in $G$ $(1\leq i\leq m)$, then $\{A_{1},\ldots,A_{m},\{g_{1}\},\ldots,\{g_{r}\}\}$ is a GSEDF in $G$ with parameters $(n,m+r;k_{1},\ldots,k_{m},\underbrace{1,\ldots,1}_{r};\lambda_{1},\ldots\lambda_{m},\underbrace{1,\ldots,1}_{r})$ where $\lambda_{i}=k_{i}-\lambda_{i}^{'}$ $(1\leq i\leq m)$.

It is well-known that if $A$ is an $(n,k,\lambda)$-DS in $G$, then, $\overline{A}=G-A$ is an $(n,n-k,n-2k+\lambda)$-DS in $G$. With this fact, the following result is a direct consequence of Theorem 2.1.

$\mathbf{Corollary\ 2.3}$ If $A$ is an $(n,k,\lambda)$-DS in $G$, then $\{A,G-A\}$ is an $(n,2;k,n-k;k-\lambda,k-\lambda)$-GSEDF in $G$.

Now we show a similar result: If $A_{1},\ldots,A_{m}$ is a partition of $G-\{0\}$, then $\{A_{1},\ldots,A_{m}\}$ is a GEDF in $G$ if and only if each $A_{i}$ $(1\leq i\leq m)$ is an $(n,k_{i},\lambda_{i},\mu_{i})$-PDS in $G$ with $\lambda_{i}=\mu_{i}-1$.

The following result is a generalization of [4], Theorem 4.6.

$\mathbf{Theoem\ 2.4}$ Let $(G,+)$ be an abelian group, $|G|=n$, $A_{1},\ldots,A_{m}$ $(m\geq2)$ be a partition of $G-\{0\}$. Then $\{A_{1},\ldots,A_{m}\}$ is an $(n,m;k_{1},\ldots,k_{m};\lambda_{1},\ldots,\lambda_{m})$-GSEDF in $G$ if and only if each $A_{i}$ $(1\leq i\leq m)$ is an $(n,k_{i},\lambda_{i}^{'},\mu_{i}^{'})$-PDS in $G$ with $\lambda_{i}^{'}=\mu_{i}^{'}-1$. In fact, $\lambda_{i}^{'}=k_{i}-\lambda_{i}-1$ and $\mu_{i}^{'}=k_{i}-\lambda_{i}$.

\pf{$\Leftarrow$:\ Suppose that $A_{i}$ is an $(n,k_{i};\lambda_{i}^{'},\mu_{i}^{'})$}-PDS in $G$ with $\lambda_{i}^{'}=\mu_{i}^{'}-1$ $(1\leq i\leq m)$. Then $|A_{i}|=k_{i}$ and

\begin{align}
\Delta (A_i,A_i)& = k_i\{0\}+\lambda^{'}_iA_i+\mu^{'}_i(G-A_{i}-\{0\})\notag\\
                                  &= k_i\{0\}+\mu^{'}_i(G-\{0\})-A_i\ (\text{since} \ \lambda^{'}_i=\mu^{'}_i-1)
\end{align}
By assumption $\sum_{j=1}^mA_i=G-\{0\}$, we get for each $i$, $1\leq i\leq m$,

\begin{align*}
\sideset{}{^{m}_{j=1, j\neq i }} \sum\Delta(A_{i},A_{j})& = \Delta(A_i, G-A_{i}-\{0\})\\
& = \Delta(A_{i},G)-\Delta(A_{i},A_{i})-\Delta(A_{i},\{0\}) \\
                                                    &= k_iG-\Delta(A_i,A_i)-A_i \ (\text{for any} \ S\subseteq G, \Delta(S,G)=|S|G)\\
                                                    &=k_i(G-\{0\})+k_i\{0\}-A_i-(k_i\{0\}+\mu_{i}^{'}
                                                    (G-\{0\})-A_i) \ (\text {by (2.1))}\\
                                                    &=(k_i-\mu_{i}^{'})(G-\{0\})
\end{align*}
Therefore $\{A_1,\ldots,A_m\}$ is an $(n,m;k_1,\ldots,k_m;\lambda_1,\ldots,\lambda_m)$-GSEDF
in $G$ where $\lambda_i=k_i-\mu_{i}^{'}=k_i-\lambda_{i}^{'}-1$ $(1\leq i\leq m).$ \\
$\Rightarrow$: Suppose that $\{A_1,\ldots,A_m\}$ is an $(n,m;k_1,\ldots,k_m;\lambda_1,\ldots,\lambda_m)$-GSEDF in $G$ and $\sum_{j=1}^{m}A_j=G-\{0\}$.
Then for each $i$ $(1\leq i\leq m)$
\begin{equation}
\sum^{m}_{j=1\atop j\neq i }\Delta(A_{i},A_{j})=\lambda_{i}(G-\{0\})
\end{equation}
The left-hand side of (2.2) is
$$\Delta(A_{i}, G-A_{i}-\{0\})=k_iG-\Delta({A_{i},A_{i}})-A_i$$
Therefore for each $i$, by (2.2)
\begin{align*}
\Delta(A_{i},A_{i})& = k_iG-A_i-\lambda_i(G-\{0\})\\
& =k_i\{0\}+(k_i-\lambda_i-1)A_i+(k_i-\lambda_i)(G-A-\{0\})
\end{align*}
which means that each $A_i$ is an $(n,k_i,\lambda_{i}^{'},\mu_{i}^{'})$-PDS in $G$ with $\mu_{i}^{'}=k_i-\lambda_i$ and $\lambda_{i}^{'}=\mu_{i}^{'}-1$. This completes the proof of
Theorem 2.4.\qed

$\mathbf{Lemma\ 2.5}$ Let $(G,+)$ be an abelian group, $|G|=n$, $A$ be a subset of $G$, $0\notin D$
and $-D=D$. If $D$ is an $(n,k;\lambda,\mu)$-PDS in $G$, then $D'=G-D-\{0\}$ is an $(n,n-k-1,n-2k+\mu-2,n-2k+\lambda)$-PDS in $G$.

\pf{Suppose that $D$ is an $(n,k,\lambda,\mu)$-PDG in $G$ which means that
\begin{align*}
\Delta(D,D)& = k\{0\}+\lambda D+\mu(G-D-\{0\})\\
& =k\{0\}+\lambda(G-D'-\{0\})+\mu D'
\end{align*}}
By assumption $-D=D$ we have $\Delta(D,\{0\})=D$, $\Delta(\{0\},D)=-D=D$ and for any $S\subseteq G$, $\Delta(G,S)=\Delta(S,G)=|S|G$. Therefore
\begin{align*}
\Delta(D',D')& = \Delta(G-\{D+\{0\}\},G-\{D+\{0\}\})\\
& = -2(k+1)G+nG+\Delta(D+\{0\},D+\{0\}) \\
&= (n-2k-2)G+\Delta(D,D)+2D+\{0\}\\
&=(n-2k-2)[\{0\}+D'+(G-D'-\{0\})]+k\{0\}+\lambda(G-D'-\{0\})+\mu D'\\
&\ \ \ +2(G-D'-\{0\})+\{0\}\\
&=(n-k-1)\{0\}+(n-2k+\mu-2)D'+(n-2k+\lambda)(G-D'-\{0\}
\end{align*}
which means that $D'$ is an $(n,n-k-1,n-2k+\mu-2,n-2k+\lambda)$-PDS in $G$.\qed

$\mathbf{Corollarly\ 2.6}$ Let $(G,+)$ be an abelian group, $|G|=n$, $A$ be an $(n,k,\lambda,\mu)$-PDS
in $G$, $\lambda=\mu-1$ and $0\notin A$, $-A=A$. Then $\{A,G-A-\{0\}\}$ is an $(n,2;k,n-k-1;k-\lambda-1,k-\lambda-1)$-GSEDF in $G$.\\
\pf{By Lemma 2.5, $D'=G-A-\{0\}$} is an $(n,n-k-1,n-2k+\mu-2,n-2k+\lambda)$-PDS in $G$. From
$\lambda=\mu-1$ we get $n-2k+\mu-2=(n-2k+\lambda)-1$. Then by Theorem 2.4 we know that the partition
$\{D, G-D-\{0\}$ of $G-\{0\}$ is an $(n,2;k,n-k-1,k-\lambda-1,k-\lambda-1)$-GSEDF in $G$.\qed

It is proved in [1] (also see [5], Theorem 13.1) that if $D$ is an $(n,k,\lambda,\mu)$-PDS in an abelian group $(G,+)$ such that $k<\frac{n}{2}$, $\lambda=\mu-1$, $0\notin D$ and $-D=D$, then the parameters should be

(\uppercase\expandafter{\romannumeral1})$(n,k,\lambda,\mu)$=$(n,\frac{n-1}{2},\frac{n-5}{4},\frac{n-1}{4})$
where $n\equiv1\pmod 4$, or

(\uppercase\expandafter{\romannumeral2})$(n,k,\lambda,\mu)$=$(243,22,1,2)$

The type (\uppercase\expandafter{\romannumeral1}) PDS is called Paley PDS. From Corollary 2.6 we know that for each Paley PDS $D$, $\{D,D^{'}\}$ is an $(n,2;\frac{n-1}{2},\frac{n+3}{4})$-SEDF in $G$ where $D^{'}=G-D-\{0\}$ (also see [4], Theorem 4.4). The first series of Paley PDS was given by quadratic cyclotomic class in finite field $\mathbb{F}_{q}$ $(q\equiv1\pmod4)$. Let $\mathbb{F}_{q}^{*}=\langle\theta\rangle$, $C_{0}=\langle\theta^{2}\rangle$ and $C_{1}=\theta C_{0}$. Then $C_{0}$ is a Paley PDS in $(\mathbb{F}_{q},+)$ with $(n,k,\lambda,\mu)=(q,\frac{q-1}{2},\frac{q-5}{4},\frac{q-1}{4})$. Thus $\{C_{0},C_{1}\}$ is a $(q,2;\frac{q-1}{2},\frac{q+3}{4})$-SEDF in $\mathbb{F}_{q}$. More general result was show by J. Polhill([8]): for any odd number $m\geq 3$, there exists an abelian group $G$ with size $m^{4}$ and $9m^{4}$ such that $G$ has a Paley PDS. On the other hand, the type (\uppercase\expandafter{\romannumeral2}) PDS has been constructed by using Golay code or finite geometry on $G=\mathbb{F}_{q}$ $(q=3^{5})$ (see {5}, example 8.3(2)). But in order to use Corollary 2.6, we need $\frac{n-1}{k}=11$ such PDS $D_{1}$, $\ldots$, $D_{11}$ and $\{D_{1},\ldots,D_{11}\}$ is a partition of $G-\{0\}$. It was shown in [10] that the cyclotomic classes $C_{i}$ $(0\leq i\leq10)$ of order $11$ in $\mathbb{F}_{q}$ $(q=3^{5})$ satisfy these condition and, by Corollary 2.6, an $(n,m,k,\lambda)=(243,11,22,20)$-SEDF in $(\mathbb{F}_{q},+)$ $(q=3^{5})$ was constructed.

In next section we will present series of GSDF's and BGSDF's in finite field $\mathbb{F}_{q}$ in more careful way by using cyclotomic classes in $\mathbb{F}_{q}$.

\section{Cyclotomic Constructions on GSDF and BGSDF}
\subsection {Cyclotomic classes and Cyclotomic Numbers}

Let $q=p^m$ be a power of prime $p$, $\mathbb{F}_q^{\ast}=\langle\theta\rangle$, $q-1=ef$ $(e\geq2)$.
Then $C=\langle\theta^e\rangle$ is the cyclic subgroup of $\mathbb{F}_q^{\ast}$ of order $f$, and the
cosets of $C$ in $\mathbb{F}_q^{\ast}$ are
$$C_\lambda=\theta^{\lambda}C\ (0\leq\theta\leq e-1),\ |C_{\lambda}|=f, \ C_0=C.$$
which are called the cyclotomic classes of order $e$ in $\mathbb{F}_{q}$.

$\mathbf{Definition \ 3.1}$ The cyclotomic numbers $(i,j)_e \ (0\leq i,j\leq e-1)$ of order $e$
on $\mathbb{F}_q$ are defined by
$$(i,j)_e=|(1+C_i)\cap C_j|=\sharp\{x\in C_i: 1+x\in C_j\}.$$
For any $l,s\in \mathbb{Z}$, we have $C_{\lambda}=C_{\lambda+le}$ and assume $(i+le,j+se)_e=(i,j)_e$.
The following properties on $(i,j)=(i,j)_e$ can be easily proved (see T. Storer's book [9]).

$\mathbf{Lemma\ 3.2}$ (1) $(i,j)=(-i,j-i),\ (i,j)=(pi,pj).$

(2) $-1\in C_{e/2}$ if $p\geq 3$ and $2\nmid f,$ and $-1\in C_0$ otherwise.

(3)\begin{displaymath}
 (i,j)
 = \left\{ \begin{array}{ll}
(j,i), & \textrm{if $-1\in C_0$ }\\
(j+\frac{e}{2},i+\frac{e}{2}), & \textrm{otherwise}

\end{array} \right.
\end{displaymath}

(4)\begin{displaymath}
 \sum_{j=0}^{e-1}(i,j)
 = \left\{ \begin{array}{ll}
f, & \textrm{if $-1\in C_i$ }\\
f-1, & \textrm{otherwise}
\end{array} \right.
\end{displaymath}

(5)$\Delta(C_{i},C_{j})=\delta_{ij}f\{0\}+\sum_{\lambda=0}^{e-1}(j-\lambda,i-\lambda)C_{\lambda}~~(0\leq i,j\leq e-1)$ where $\delta_{ij}=1$ if $i=j$ and $\delta_{ij}=0$ otherwise.\qed

The cyclotomic numbers were determined for several smaller order $e$ and semiprimitive case
in [9] and [3]. Now we use parts of computed values of cycltomic numbers and the formula in Lemma 3.2 (5) to show that some cyclotomic classes in finite field give GSEDF or BGSEDF.
\subsection{For small $e$ cases}
\subsubsection{$e=2$}

Let $q=p^{m}$, $p\geq3$, $q-1=2f$, $(i,j)=(i,j)_{2}$ $(0\leq i,j\leq 1)$. Let $\mathbb{F}^{*}_{q}=\langle\theta\rangle$, $C_{0}=\langle\theta^{2}\rangle$ and $C_{1}=\theta C_{0}$ be the quadratic cyclotomic classes in $\mathbb{F}_{q}$. Then (see[9])

For $q\equiv 3 ~(mod~4)$, $(0,0)=(1,1)=(1,0)=\frac{q-3}{4}$, $(0,1)=\frac{q+1}{4}$. From Lemma 3.2(5) we know that
\begin{align*}
\Delta (C_i,C_i)& = \frac{q-1}{2}\{0\}+(i,i)C_0+(i+1,i+1)C_1 \\
& =\frac{q-1}{2}\{0\}+\frac{q-3}{4}(C_0+C_1)\\
& =\frac{q-1}{2}\{0\}+\frac{q-3}{4}(\mathbb{F}_q-\{0\})) \ (i=1,2)
\end{align*}
which means that $C_i(i=0,1)$ is an $(n,k,\lambda)=(q,\frac{q-1}{2},\frac{q-3}{4}))$-DS
in $\mathbb{F}_q$.

For $q\equiv 1\bmod4,$ $-1\in C_0$ and $(0,1)=(1,0)=(1,1)=\frac{q-1}{4}$, $(0,0)=\frac{q-5}{4}$. Therefore
\begin{align*}
\Delta (C_0,C_0)& = \frac{q-1}{2}\{0\}+(0,0)C_0+(1,1)C_1 \\
& =\frac{q-1}{2}\{0\}+\frac{q-5}{4}C_0+\frac{q-1}{4}(\mathbb{F}_q-C_0-\{0\})
\end{align*}
which means that $C_0$ is an $(n,k,\lambda,\mu)=(q,\frac{q-1}{2},\frac{q-5}{4},\frac{q-1}{4})$-PDS in
$\mathbb{F}_q$. By Corollary 2.3 and 2.6 we get the following results

$\mathbf{Theorem 3.3}$ ~ Let $q=p^m$, $q-1=2f$, $\mathbb{F}_q^{\ast}=\langle\theta\rangle$, $C_0=\langle\theta^2\rangle$, $C_1=\theta C_0$.

(1) ([4]) If $q\equiv 1(\bmod 4),$ $\{C_0,C_1\}$ is a $(q,2;\frac{q-1}{2},\frac{q-1}{4})$-SEDF in $\mathbb{F}_q$.

(2) If $q\equiv 3(\bmod 4)$, $\{C_0,C_1+\{0\}\}$ and $\{C_1,C_0+\{0\}\}$ are $(q,2;\frac{q-1}{2},\frac{q+1}{2},\frac{q+1}{4},\frac{q+1}{4})$-GSEDF in $\mathbb{F}_q$.

\subsubsection{$e=4$}

Let $q=p^m\equiv1(\bmod4)$, $q-1=4f$, $\mathbb{F}_q^{\ast}=\langle\theta\rangle$, $C_{\lambda}=\theta^{\lambda}\langle\theta^4\rangle$ $(0\leq\lambda\leq 3)$.

If $q=1+16t^2$, it is proved that $\{C_0,C_2\}$ (and $\{C_1,C_3\}$) is $(q,\frac{q-1}{4},\frac{q-1}{16})$-SEDF
in $\F_q$ ([2], Theorem 4.3).

It is proved in [9] that if $q=1+4t^2\ (2\nmid t)$, then $C_i (0\leq i\leq 3)$ is a $(q,\frac{q-1}{4},\frac{q-5}{16})$-DS
in $\F_q$, and if $q=9+4t^2 (2\nmid t), C_i+\{0\} (0\leq i\leq 3)$ is a $(q,\frac{q+3}{4},\frac{q+3}{16})$-DS in $\F_q$.
By Corollary 2.2 and 2.3 we get

\textbf{Theorem 3.4}~  Let $q=p^m\equiv 1\pmod 4, C_i (0\leq i\leq 3)$ be the cyclotomic classes of order 4 in $\F_q$. Then

(1) If $q=1+4t^2 (2\nmid t)$, for any $i (0\leq i\leq 3)$, $\{C_i,\F_q-C_i \}$ is a
$(q,2;\frac{q-1}{4},\frac{3q+1}{4};\frac{3q+1}{16},\frac{3q+1}{16})$-GSEDF in $\F_q$ and $\{C_0,C_1,C_2,C_3,\{0\}\}$
is a $(q,5;\frac{q-1}{4},\frac{q-1}{4},\frac{q-1}{4},\frac{q-1}{4},1;\frac{3q-3}{16},\frac{3q-3}{16},\frac{3q-3}{16},\frac{3q-3}{16},1)$-GSEDF in $\F_q$.

(2) If $q=9+4t^2 (2\nmid t)$, for any $i (0\leq i\leq 3)$, $\{C_i+\{0\},\F_q-C_i-\{0\}\}$ is a
$(q,2;\frac{q+3}{4},\frac{3q-3}{4};\frac{3q+9}{16},\frac{3q+9}{16})$-GSEDF in $\F_q$.

\subsubsection{$e=6$}

Let $q=p^m\equiv 1\pmod 6, C_{\lambda} (0\leq \lambda\leq 5)$ be the cyclotomic classes of order 6 in $\F_q$.

If $q=1+108t^2$, it is proved that $\{C_0,C_3\}$ (and $\{C_1,C_4\}, \{C_2,C_5\}$) is a
$(q,2;\frac{q-1}{6},\frac{q-1}{36})$-SEDF in $\F_q$ ([2], Theorem 4.6, where q is a prime, but it
works for q being a power of prime).

\subsubsection{$e=8$}

Let $q=p^m\equiv 1\pmod 8, C_{\lambda} (0\leq \lambda \leq 7)$ be the cyclotomic classes of order 8 in $\F_q, (i,j)=(i,j)_8 (0\leq i,j\leq 7)$
be the cyclotomic numbers of order 8 on $\F_q$. It is proved in [9] that

(1) ([9] Part 1, Theorem 19) If $q=9+64y^2=1+8b^2,y,b\in\Z$, and $2\nmid y$, then for each $i\ (0\leq i\leq 7)$,
$C_i$ is a $(q,\frac{q-1}{8},\frac{q-9}{64})$-DS in $\F_q$.
The first two examples are $q=73$ and $140411704393$.

(2) ([9] Part 1, Theorem $9^{'}$) If $q=441+64y^2=49+8b^2$ and $y,b\in\Z$, then for each $i\ (0\leq i\leq 7)$,
$C_i+\{0\}$ is a  $(q,\frac{q+7}{8},\frac{q+7}{64})$-DS in $\F_q$. The first example is $q=26041$.

By Corollary 2.2 and 2.3 we get

\textbf{Theorem 3.5} Let $q=p^m\equiv 1\pmod 8$ and $C_{\lambda} (0\leq \lambda \leq 7)$ be the cyclotomic
classes of order 8 in $\F_q$.

(1) If $q=9+64y^2=1+8b^2,y,b\in\Z$, and $2\nmid y$, then for each $i\ (0\leq i\leq 7)$,
$\{C_i,\F_q-C_i\}$ is a $(q,2;\frac{q-1}{8},\frac{7q+1}{8};\frac{7q+1}{64},\frac{7q+1}{64})$-GSEDF in $\F_q$,
and $\{C_0,C_1,\ldots,C_7,\{ 0\}\}$ is a $(q,9;\underbrace{\frac{q-1}{8},\ldots,\frac{q-1}{8}}_{8},1;
\underbrace{\frac{7q+1}{64},\ldots,\frac{7q+1}{64}}_{8},1)$-GSEDF in $\F_q$.

(2) If $q=441+64y^2=49+8b^2$ and $y,b\in\Z$, then for each $i\ (0\leq i\leq 7)$,
$\{C_i+\{0\},\F_q-C_i-\{0\}\}$ is a  $(q,2;\frac{q+7}{8},\frac{7(q-1)}{8};\frac{7(q+7)}{64},\frac{7(q+7)}{64})$-GSEDF in $\F_q$.

\subsection{Semiprimitive case}
Let $q=p^m$ and $q-1=ef (e\geq 2)$. The following condition is called semiprimitive

$(\ast)$ there exists $t\in \mathbb{Z}$ such that $p^{t}\equiv-1~(mod~e)$.

In this case, the cyclotomic numbers $(i,j)=(i,j)_e\ (0\leq i,j\leq e-1)$ in $\F_q$
have been given in [3].

\textbf{Lemma 3.6} ([3], also see [5], Theorem 10.3)
Let $q=p^m$ and $q-1=ef (e\geq 2)$. Assume that the semiprimitive condition ($\ast$) holds. Then $m=2l$,
and there exists $\varepsilon\in\{\pm 1 \}$ such that $s=\varepsilon p^l=\varepsilon \sqrt{q}\equiv 1\pmod e$.
Let $(i,j)=(i,j)_e\ (0\leq i,j\leq e-1)$ be the cyclotomic numbers of order $e$ in $\F_q$. Then
\[
\begin{aligned}
(0,0)&=\eta^2-(e-3)\eta-1,\ (0,i)=(i,0)=(i,i)=\eta^2+\eta\ (1\leq i\leq e-1) \\
(i,j)&=\eta^2 (1\leq i\neq j\leq e-1),
\end{aligned}
\]
where $\eta=\frac{s-1}{e}$.

The following result shows that Lemma 3.6 can be used to construct BGSEDF. Recall Definition 1.1 of BGSEDF:
for subset $A_1,\ldots,A_m (m\geq 2)$ of $\F_q$, $\{A_1,\ldots,A_m\}$ is call a
$(q,m;k_1,\ldots,k_m;\lambda_1,\ldots,\lambda_m)$-BGSEDF in $\F_q$ if for each $i\ (1\leq i\leq m)$
\[
\Delta(A_i,\sum_{j=1\atop j\neq i}^{m}A_j)\leq \lambda_i(\F_q-\{0\}).
\]

The BGSEDF with $\lambda_1=\cdots=\lambda_m=1$ is more interesting since, as mentioned in Section 1,
such BGSEDF is equivalent to G-optimal strong AMD mode.

\textbf{Theorem 3.7} Let $q=p^{2l}, l\geq 1, q-1=ef,e=p^l+1=\sqrt{q}+1,f=\sqrt{q}-1, C_{\lambda}\ (0\leq \lambda \leq e-1)$
be the cyclotomic classes of order $e$ in $\F_q$. Suppose $i$ and $j$ are distinct integers in $\{0,1,\ldots,e-1 \}$
and $S_i,S_j$ are subsets of $C_i$ and $C_j$ respectively, $|S_i|=k_i\geq 1,|S_j|=k_j\geq 1$.
Then $\{S_i,S_j\}$ is a $(q,2;k_i,k_j;1,1)$-BGSEDF and $\{S_i+\{0\},S_j\}$ is a $(q,2;k_i+1,k_j;1,1)$-BGSEDF
in $\F_q$.

\pf{
From $p^l\equiv -1\pmod e$ we know that the semiprimitive condition ($\ast$) holds. Using Lemma 3.6, we have
$s=-\sqrt{q}$ and $\eta=\frac{s-1}{e}=-1$. Therefore
\[
(0,i)=(i,0)=(i,i)=\eta^2+\eta=0\ (1\leq i\leq e-1),\ (i,j)=\eta^2=1\ (1\leq i\neq j\leq e-1)
\]
where $(i,j)=(i,j)_e$. Then
\[
\Delta(S_i,S_j)\leq \Delta(C_i,C_j)=\sum_{\lambda=0}^{e-1}(i-\lambda,j-\lambda)C_{\lambda}.
\]
From assumption $0\leq i\neq j\leq e-1$ we know that for any $\lambda, i-\lambda \not\equiv 0\pmod e$
or $j-\lambda\not\equiv 0\pmod e$, and then $(i-\lambda,j-\lambda)=0$ or $1$. Therefore
$\Delta(S_i,S_j)\leq \sum_{\lambda=0}^{e-1}C_{\lambda}=\F_q-\{0\}$.
 Thus $\{S_i,S_j\}$ is a $(q,2;k_i,k_j;1,1)$-BGSEDF
in $\F_q$. Moreover, it is easy to see that $-1\in C_0$ and $-C_{\lambda}=C_{\lambda} (0\leq \lambda\leq e-1)$.
We have
\[
\begin{aligned}
\Delta(S_i+\{0\},S_j)&\leq \Delta(C_i+\{0\},C_j) \\
                     &=\Delta(C_i+C_j)+C_j=\sideset{}{_{\lambda=0,\lambda\neq j}^{e-1}}\sum(i-\lambda,j-\lambda)C_{\lambda}+((i-j,0)+1)C_j \\
                     &\leq \sideset{}{_{\lambda=0}^{e-1}}\sum C_{\lambda}=\F_q-\{0\} \  \ (\text{since}\ (i-j,0)=0).
\end{aligned}
\]
Therefore $\{S_i+\{0\},S_j\}$ is a
$(q,2;k_i+1,k_j;1,1)$-BGSEDF in $\F_q$. \qed
}

\section{Conclusion}
In this paper we present several cyclotomic constructions of SEDF, GSEDF and BSEDF in finite fields,
based on two general results (Theorem 2.1 and Theorem 2.4) which say that a family
$\{A_1,\ldots,A_m\}$ of subsets of an abelian group $(G,+)$ is a GSEDF in $G$ if the family is a partition
of $G$ and each $A_i$ is a DS in $G$, or the family is a partition of $G-\{0\}$ and each $A_i$
is an $(n,k;\lambda_i,\mu_i)$-PDS in $G$ with $\lambda_i=\mu_i-1$. All SEDF's constructed
in this paper are ones having $m=2$. The problem on existence of SEDF with $m\geq 5$ and $m\neq 11$
is still open. Another interesting problem is to develop other methods on constructing
GSEDF and BGSEDF in arbitrary finite abelian group G.

\end{document}